# A generalized expression for filling congruent circles in a circle

Ajeet K. Srivastav

*Department of Metallurgical and Materials Engineering, Visvesvaraya National Institute of Technology, Nagpur-440010, India.*

**Correspondence details**: Tel: +91−712−1351, Email: srivastav.ajeet.kumar@gmail.com; ajeet.srivastav@mme.vnit.ac.in

**Abstract**

The paper reports a generalized expression for filling the congruent circles (of radius *r*) in a circle (of radius *R*). First, a generalized expression for the biggest circle (*r*) inscribed in the $n^{th}$ part of the bigger circle (*R*) was developed. Further, it was extended as *n* such circles (*r*) touching each other and the bigger circle (*R*). To fill the bigger circle (*R*), the exercise was further repeated by considering the bigger circle radius as *R*-2*r*, *R*-4*r* and so on. In the process, a generalized expression was deduced for the total no. of such circles (*r*) which could be inscribed in this way of filling the bigger circle (*R*). The approach does not claim the closest packing always though it could be helpful for practical purposes.

**Keywords:** Congruent Circles; Filling; Circle; Analytical approach

## 1. Introduction

Filling circles in bounded shapes have been an intriguing problem since long in pure mathematics literature. Specifically, filling non-overlapping identical circles in a circle has received tremendous interest considering the fundamental insights and the industrial applications to avoid the wastage of materials [1, 15]. Also, Tarnai and Miyazaki [14] reported the interesting connection between filling circles in a circle, lotus receptacles, and the ancient Japanese art.

Kravitz was one of the first to contribute to the problem of filling circles in a circle [9]. Over the period, several contributions have been made towards the problem. However, the approaches were mostly focused on optimization and further proof [2-8, 10-12] due to the complexity of the analytical approaches. The topical website http://www.packomania.com/ [13] should be referred for the best-known solutions and the related references.

In this paper, the filling of congruent circles in a circle has been considered through an analytical approach. First, a generalized expression for the biggest possible circle in the $n^{th}$ part of a circle was derived. Subsequently, the approach was extended considering the concentric circles inside the starting circle to develop the generalized expression for the total no. of circles inscribed in the circle. Further, the working principle has been demonstrated with examples. To the best of author's knowledge, no generalized expression is available for the same in the literature.

**2. Biggest circle radius (*r*) in $n^{th}$ part of the circle (*R*)**

First, a generalized expression for the biggest circle radius inscribed in the $n^{th}$ part of a circle was developed. Here, we are interested to inscribe the biggest circle with radius *r* inside the $n^{th}$ part of the bigger circle of radius *R*. The biggest circle is the one which finally touches all the boundaries of the part of the circle. The first idea is to start with a small circle touching two boundaries and let it grow until it touches the third boundary to get the biggest circle. If we start with any of the two boundaries including arc, eventually one of the boundaries will be crossed in due process of the growth of the circle as both the boundaries move unevenly due to symmetrical geometry of the circle. The only logical option is to choose two boundaries that move evenly in both directions. Fig. 1 shows the result for the biggest circle (*r*) inscribed in the $n^{th}$ part of the circle (*R*) which makes angle $2\theta$ at the center.

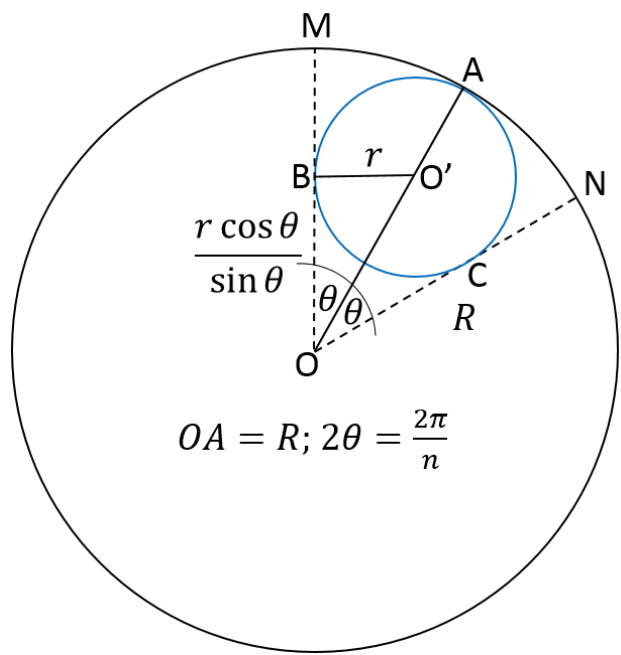

**Fig. 1.** Schematic for the biggest circle (*r*) inscribed in the $n^{th}$ part of the bigger circle (*R*).

From Fig.1, we can write $OA - OO' = O'A$, i.e.

$$R - \frac{r}{\sin\theta} = r \text{ or } r = \frac{R}{\left(1 + \frac{1}{\sin\theta}\right)} \quad (1)$$

We know that $2\theta = \frac{2\pi}{n}$, therefore the generalized expression for the radius of the biggest possible circle ($r$) inscribed in the $n^{th}$ part of the bigger circle ($R$) is

$$r = \frac{R}{\left(1 + \frac{1}{\sin\left(\frac{2\pi}{2n}\right)}\right)}; n \neq 1 \quad (2)$$

We can estimate the biggest circle radius by giving different $n$ values.

*Examples*

**1.** For $n = 2; r = \dfrac{R}{\left(1 + \dfrac{1}{\sin\left(\dfrac{2\pi}{2\times 2}\right)}\right)} = \dfrac{R}{2}$

**2.** For $n = 3; r = \dfrac{R}{\left(1 + \dfrac{1}{\sin\left(\dfrac{2\pi}{2\times 3}\right)}\right)} = \dfrac{R}{1 + \dfrac{2}{\sqrt{3}}} = \dfrac{\sqrt{3}R}{\left(2 + \sqrt{3}\right)}$

**3.** For $n = 4; r = \dfrac{R}{\left(1 + \dfrac{1}{\sin\left(\dfrac{2\pi}{2\times 4}\right)}\right)} = \dfrac{R}{\sqrt{2} + 1}$, and so on.

The radius of the inscribed circle ($r$) with respect to the bigger circle ($R$) could be represented as $r/R$ which varies with $n$ as illustrated in Fig. 2. The graph was plotted up to $n = 180$ divisions.

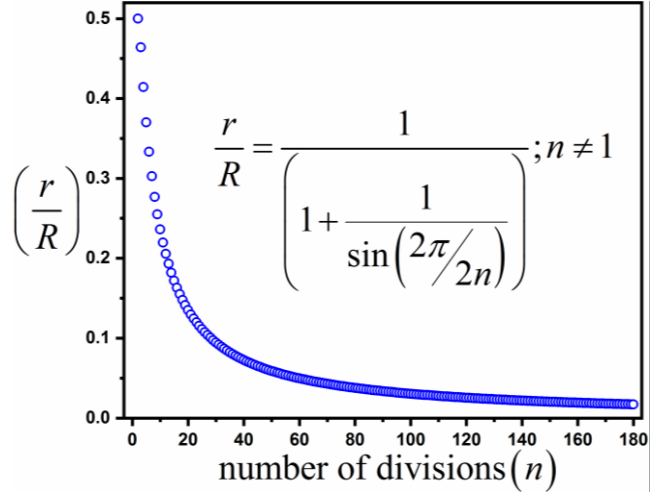

**Fig. 2.** The radius of the biggest inscribed circle (*r*) with respect to the radius of the bigger circle (*R*), (*r/R*) variation with the no. of divisions (*n*).

**3. A generalized expression for the total no. (*N*) of circles (*r*) in the bigger circle (*R*)**

The whole approach of filling circles is to divide the bigger circle (*R*) in concentric circles of radius *R*, *R-2r*, *R-4r*, *R-6r*, and so on. Further, all these concentric circles will be filled by the congruent circles without overlapping each other by the approach and the formula obtained in the previous section. It is worthy to point out here that the no. of circles in each row is going to be the greatest positive integer number less than or equal to the calculated number. The no. of circles possible in the starting circle from equation 2 is:

$$n_1 = \left\lfloor \frac{\pi}{\sin^{-1}\left(\frac{r}{R-r}\right)} \right\rfloor \tag{3}$$

The $\lfloor \; \rfloor$ refers to the greatest integer less than or equal to a number [8].

Accordingly, filling of the same congruent circles touching the second, third, and further up to $m^{th}$, concentric circles could be represented as:

$$n_2 = \left\lfloor \frac{\pi}{\sin^{-1}\left(\frac{r}{R-3r}\right)} \right\rfloor ; n_3 = \left\lfloor \frac{\pi}{\sin^{-1}\left(\frac{r}{R-5r}\right)} \right\rfloor ;$$

$$\ldots\ldots; n_{k-1} = \left\lfloor \frac{\pi}{\sin^{-1}\left(\frac{r}{R-(2k-3)r}\right)} \right\rfloor ; n_k = \left\lfloor \frac{\pi}{\sin^{-1}\left(\frac{r}{R-(2k-1)r}\right)} \right\rfloor ; \quad (4)$$

$$\ldots\ldots; n_{m-1} = \left\lfloor \frac{\pi}{\sin^{-1}\left(\frac{r}{R-(2m-3)r}\right)} \right\rfloor ; n_m = \left\lfloor \frac{\pi}{\sin^{-1}\left(\frac{r}{R-(2m-1)r}\right)} \right\rfloor$$

Therefore, the total no. of such identical circles filled in the starting circle ($R$) is:

$$N = n_1 + n_2 + n_3 + \ldots + n_{k-1} + n_k + \ldots + n_{m-1} + n_m = \sum_{k=1}^{m} n_k \quad (5)$$

Here, $m$ is the total number of possible concentric circles including the starting circle as well. It is worthy to point out here that the radius of the $m^{th}$ concentric circle is $R-(2m-2)r$. The schematic of the possible configurations based on $r$ and $R$ is shown in Fig. 3. In case I, the $m^{th}$ row filling is not possible as the corresponding concentric circle radius $R-(2m-2)r$ is smaller than the filling circle ($r$). In case II and case III, the $m^{th}$ concentric circle is possible as the radius $R-(2m-2)r$ is equal or more than the filling circle radius ($r$). However, the filling is not allowed in the present approach. Moreover, one can consider putting one extra filling circle ($r$) concentric to the starting circle ($R$). In case IV and case V, both the drawing of $m^{th}$ concentric circle and further filling of circles ($r$) are possible as the radius of the $m^{th}$ concentric circle $R-(2m-2)r$ is the same or greater than the diameter of the filling circle ($2r$).

To summarize the possible configurations based on possible concentric circles and further practicing the filling approach, we can have three conditions possible as $R-(2m-1)r \geq r$, $0 \leq R-(2m-1)r < r$, and $R-(2m-1)r < 0$.

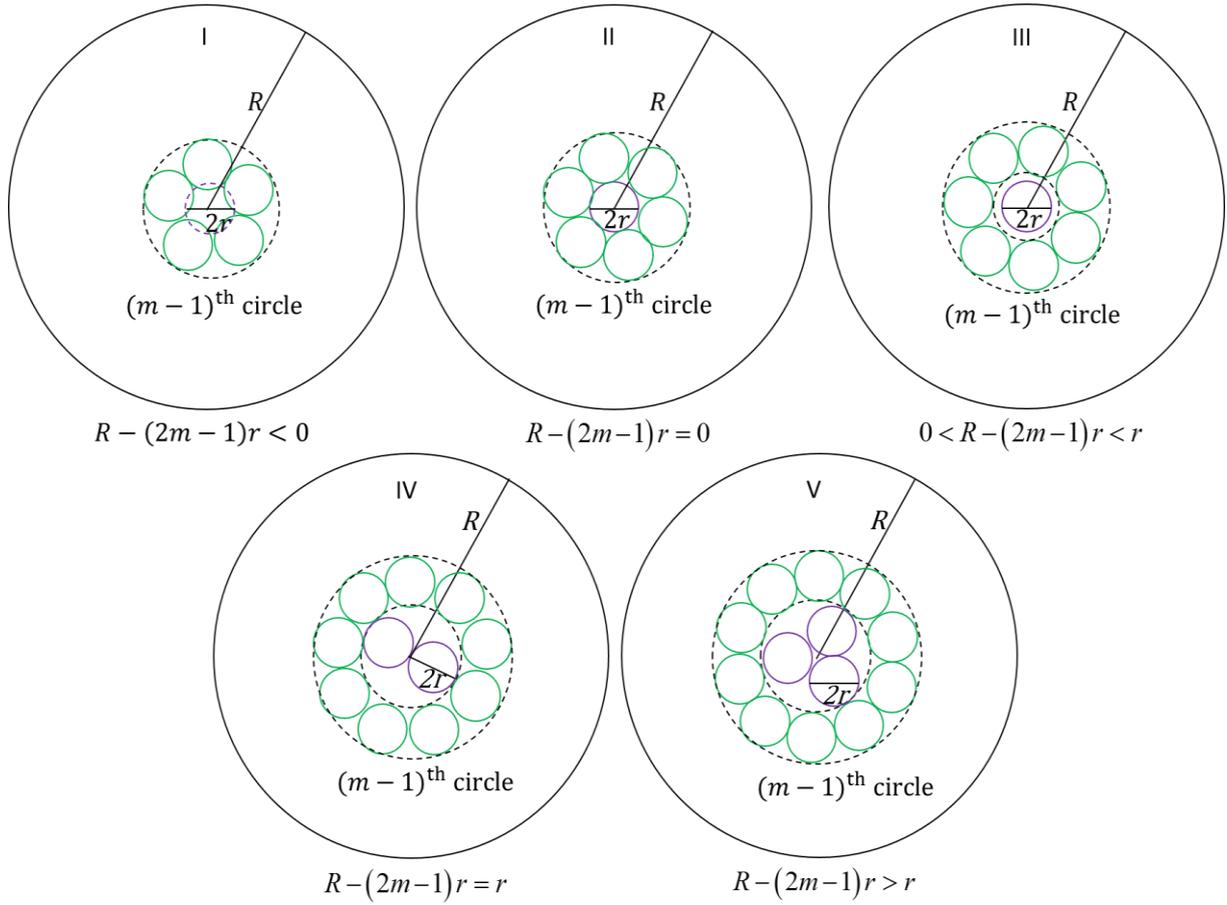

**Fig. 3.** The schematic of the possible configurations based on *r* and *R*. The dotted circles are the $m^{th}$ and $(m-1)^{th}$ concentric circles considered to practice the filling approach.

Finally, based on *r* and *R*, up to what $m^{th}$ term we are going to have in the final calculation of total no. of circles (*N*), we can write a generalized expression as:

$$N = \sum_{k=1}^{m} \left\lfloor \frac{\pi}{\sin^{-1}\left(\frac{r}{R-(2k-1)r}\right)} \right\rfloor ; R-(2m-1)r \geq r \dots\dots\dots(a)$$

$$N = \left(\sum_{k=1}^{m-1} \left\lfloor \frac{\pi}{\sin^{-1}\left(\frac{r}{R-(2k-1)r}\right)} \right\rfloor\right) + 1 ; 0 \leq R-(2m-1)r < r \dots\dots(b) \quad (6)$$

$$N = \sum_{k=1}^{m-1} \left\lfloor \frac{\pi}{\sin^{-1}\left(\frac{r}{R-(2k-1)r}\right)} \right\rfloor ; R-(2m-1)r < 0 \dots\dots\dots(c)$$

If we consider the starting circle as a *unit* circle in which the identical circles (of radius $x = \frac{r}{R}$) to be filled up, the expression transforms to:

$$N = \sum_{k=1}^{m} \left\lfloor \frac{\pi}{\sin^{-1}\left(\frac{x}{1-(2k-1)x}\right)} \right\rfloor ; 1-(2m-1)x \geq x \quad \text{...............(a)}$$

$$N = \left(\sum_{k=1}^{m-1} \left\lfloor \frac{\pi}{\sin^{-1}\left(\frac{x}{1-(2k-1)x}\right)} \right\rfloor\right) + 1; 0 \leq 1-(2m-1)x < x \quad \text{...........(b)} \qquad (7)$$

$$N = \sum_{k=1}^{m-1} \left\lfloor \frac{\pi}{\sin^{-1}\left(\frac{x}{1-(2k-1)x}\right)} \right\rfloor ; 1-(2m-1)x < 0 \quad \text{...............(c)}$$

*Examples*

**1.** Taking 1$^{st}$ case for the demonstration purpose here as $R = 5$ cm, and $r = 1.67$ cm; $\frac{r}{R}$ or $x = 0.334$, the condition $R-(2m-1)r = 0$; or $1-(2m-1)x = 0$ fulfills here for $m = 2$. Therefore, equation 7(b) is applicable here and the summation is going to expand just for the 1$^{st}$ term only. So, from equation 7(b), we have

$$N = \left(\sum_{k=1}^{m-1} \left\lfloor \frac{\pi}{\sin^{-1}\left(\frac{x}{1-(2k-1)x}\right)} \right\rfloor\right) + 1 = \left(\sum_{k=1}^{1} \left\lfloor \frac{\pi}{\sin^{-1}\left(\frac{x}{1-(2k-1)x}\right)} \right\rfloor\right) + 1 = \left\lfloor \frac{\pi}{\sin^{-1}\left(\frac{x}{1-x}\right)} \right\rfloor + 1$$

$$= \left\lfloor \frac{\pi}{\sin^{-1}\left(\frac{0.334}{1-0.334}\right)} \right\rfloor + 1 = \left\lfloor \frac{\pi}{\sin^{-1}(0.5)} \right\rfloor + 1 = \left\lfloor \frac{\pi}{\left(\frac{\pi}{6}\right)} \right\rfloor + 1 = 07$$

The example is illustrated in Fig. 4. This is the best possible filling configuration for 07 number of circles in a circle which was proven by R.L. Graham in 1968 [6].

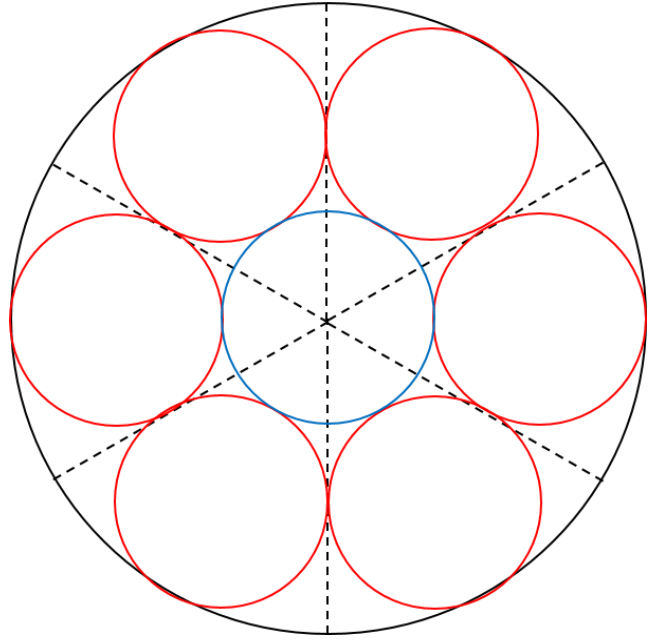

**Fig. 4.** The approach is demonstrated taking $R = 5$ cm, and $r = 1.67$ cm; $\frac{r}{R}$ or $x = 0.334$. The total number of circles are seven (07).

**2.** Taking another case for the demonstration purpose here as $R = 4.8$ cm, and $r = 0.6$ cm; $\frac{r}{R}$ or $x = 0.125$, the condition $R - (2m-1)r = r$; or $1 - (2m-1)x = x$ fulfills here for $m = 4$. Therefore, equation 7(a) is applicable here and the summation is going to expand up to 4$^{th}$ term only. So, from equation 7(a), we have

$$N = \sum_{k=1}^{m} \left\lfloor \frac{\pi}{\sin^{-1}\left(\frac{x}{1-(2k-1)x}\right)} \right\rfloor = \sum_{k=1}^{4} \left\lfloor \frac{\pi}{\sin^{-1}\left(\frac{x}{1-(2k-1)x}\right)} \right\rfloor$$

$$= \left\lfloor \frac{\pi}{\sin^{-1}\left(\frac{0.125}{1-0.125}\right)} \right\rfloor + \left\lfloor \frac{\pi}{\sin^{-1}\left(\frac{0.125}{1-0.375}\right)} \right\rfloor + \left\lfloor \frac{\pi}{\sin^{-1}\left(\frac{0.125}{1-0.625}\right)} \right\rfloor + \left\lfloor \frac{\pi}{\sin^{-1}\left(\frac{0.125}{1-0.875}\right)} \right\rfloor$$

$$= 21 + 15 + 09 + 02 = 47$$

The filling of the circles is shown in Fig. 5. One peculiarity is observed here that each row is not necessarily filled up completely. This is due to the reason of getting the non-integer positive value calculated for that row. And, we are choosing the greatest positive integer less than or equal to the calculated number. The filling density can be calculated based on the number of circles (*r*) filled in the bigger circle (*R*). The calculated density of filling here is 0.734375. The highest density for filling the same no. of circles $(N = 47)$ is 0.787760 [7, 13]. This clearly shows that the approach does not guarantee the highest filling density configuration. However, there could be a possibility for filling the same number of circles with a slightly larger radius which might further improve the filling density.

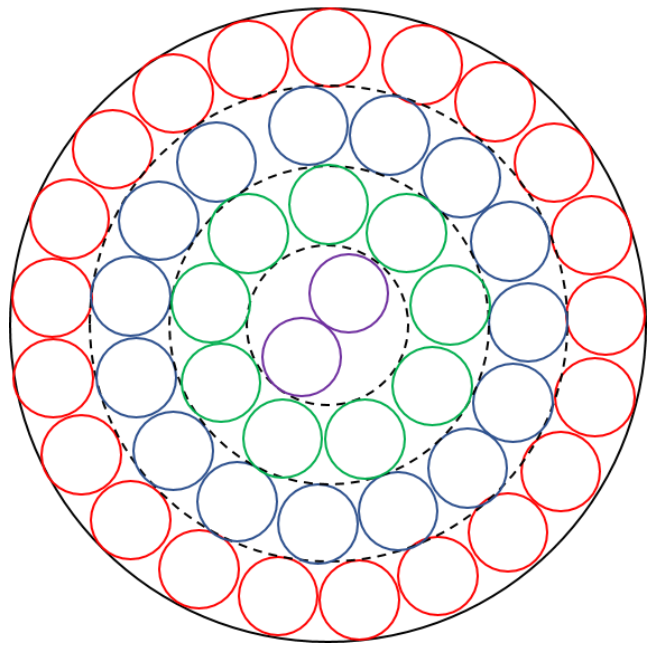

**Fig. 5.** The approach is demonstrated taking $R = 4.8$ cm, and $r = 0.6$ cm; $\frac{r}{R}$ or $x = 0.125$. The total number of circles are forty-seven (47).

## 4. Conclusions

A generalized expression for the radius of the biggest circle (*r*) inscribed in the $n^{th}$ part of the circle (*R*) was developed. The approach was furthered to achieve a generalized expression for the filling of congruent circles (*r*) in a circle (*R*). To the best of author's knowledge, there is no such analytical approach and expression available in the literature. The results are important due to their fundamental nature and could be helpful further for practical applications of filling circles in a circle. However, it should be noted here that formalism does not go always with the closest packing.

**Acknowledgments**

The author gratefully acknowledges the inputs of Jitendra Itankar in designing the problem statement and fruitful discussion with Suresh Bandi and Prashant Kumar Singh.

**References**

[1] Castillo, I., Kampas, F. J., Pintér, J. D. (2008). Solving circle packing problems by global optimization: Numerical results and industrial applications. *Eur. J. Oper. Res.* 191(3): 786–802. DOI: 10.1016/j.ejor.2007.01.054

[2] Fodor, F. (1999). The densest packing of 19 congruent circles in a circle. *Geom. Dedicata* 74(2): 139–145. DOI: 10.1023/A:1005091317243

[3] Fodor, F. (2000). The densest packing of 12 congruent circles in a circle. *Beiträge Zur Algebr. Und Geom.* 41(2): 401–409. http://eudml.org/doc/231828

[4] Fodor, F. (2003). The densest packing of 13 congruent circles in a circle. *Beiträge Zur Algebr. Und Geom.* 44(2): 431–440. http://eudml.org/doc/123619

[5] Goldberg, M. (1971). Packing of 14, 16, 17 and 20 Circles in a Circle. *Math. Mag.* 44(3): 134–139. DOI: 10.1080/0025570X.1971.11976122

[6] Graham, R. L. (1968). Sets of points with given minimum separation (Solution to Problem E1921). *Am. Math. Mon.* 75(2): 192–193. DOI: 10.1080/00029890.1968.11970965

[7] Graham, R. L., Lubachevsky, B. D., Nurmela, K. J., Ostergard, P. R. J. (1998). Dense packings of congruent circles in a circle. *Discrete Mathematics*. 181(1–3): 139–154. DOI: 10.1016/S0012-365X(97)00050-2

[8] Graham, R. L., Knuth, D. E., Patashnik, O. (1994). Integer Functions. In: *Concrete Mathematics: A Foundation for Computer Science*. 2nd edn. Reading, MA: Addison-Wesley, pp. 67–101.

[9] Kravitz, S. (1967). Packing Cylinders into Cylindrical Containers. *Math. Mag.* 40(2): 65–71. DOI: 10.1080/0025570X.1967.11975768

[10]  Melissen, H. (1994). Densest packings of eleven congruent circles in a circle. *Geom. Dedicata* 50(1): 15–25. DOI: 10.1007/BF01263647

[11]  Pirl, U. (1969). Der Mindestabstand vonn in der Einheitskreisscheibe gelegenen Punkten. *Math. Nachrichten* 40(1–3): 111–124. DOI: 10.1002/mana.19690400110

[12]  Reis, G. E. (1975). Dense Packing of Equal Circles within a Circle. *Math. Mag.* 48(1): 33–37. DOI: 10.1080/0025570X.1975.11976434

[13]  Specht, E. (2014). The best known packings of equal circles in a circle. Last update, August 12. Accessed: 4 April 2020. Available at: http://www.packomania.com/

[14]  Tarnai, T., Miyazaki, K. (2003). Circle Packings and the Sacred Lotus. *Leonardo* 36(2): 145–150. DOI: 10.1162/002409403321554215

[15]  Zhang, Z., Wu, W., Ding, L., Liu, Q., Wu, L. (2013). Packing Circles in Circles and Applications. In: Pardalos, P., Du, D. Z., Graham, R., eds. *Handbook of Combinatorial Optimization*, Vol. 4, 2nd ed. New York, NY: Springer New York, pp. 2549–2584.